\documentclass[
reprint,
amsmath,amssymb,
aps,
prb,
]{revtex4-2}

\usepackage{hyperref}
\hypersetup{colorlinks=true,
            linkcolor=black,
            citecolor=black,
            urlcolor=blue}
\usepackage{amssymb}
\usepackage{mathbbol}
\usepackage{courier}

\usepackage{graphicx}
\usepackage{dcolumn}
\usepackage{bm}

\newcommand{\rev}[1]{\textcolor{black}{#1}}

\begin{document}

\title{BCS surrogate models for floating superconductor-semiconductor hybrids}

\author{Virgil V. Baran$^{1,2,3}$}
\email[]{virgil.v.baran@unibuc.ro}
\author{Jens Paaske$^{3}$}
\affiliation{$^{1}$Faculty of Physics, University of Bucharest, 405 Atomi\c stilor, RO-077125, Bucharest-M\u agurele, Romania}
\affiliation{$^{2}$``Horia Hulubei" National Institute of Physics and Nuclear Engineering, 30 Reactorului, RO-077125, Bucharest-M\u agurele, Romania}
\affiliation{$^{3}$Center for Quantum Devices, Niels Bohr Institute, University of Copenhagen, 2100 Copenhagen, Denmark}

\begin{abstract}
Superconductor-semiconductor hybrid devices, involving quantum dots interfaced with floating and/or grounded superconductors, have reached a level of complexity which calls for the development of versatile and numerically efficient modelling tools. Here, we propose an extension of the surrogate model solver for sub-gap states [Phys. Rev. B 108, L220506 (2023)], which is able to handle floating superconducting islands with finite charging energy. Upon eliminating all finite-size effects of the computationally demanding Richardson model approach, we achieve a more efficient way of calculating the sub-gap spectra and related observables without compromising their accuracy. We provide a number of benchmarks between the two approaches and showcase the versatility of the extended surrogate model solver by studying the stability of spin-triplet ground states in various tunable devices. The methods introduced here set the stage for reliable microscopic simulations of complex superconducting quantum circuits across all their relevant parameter regimes.


\end{abstract}

\maketitle

\section{Introduction}

Hybrid superconductor-semiconductor devices combine the two key qualities of both materials: macroscopic quantum coherence and electrical tunability~\cite{DeFranceschi2010Oct, Aguado2020Dec}. This makes them highly valuable for quantum technological applications like gatemons~\cite{Casparis2018Oct} and parametric amplifiers~\cite{Phan2023Jun, Splitthoff2024Jan}, and they provide a promising platform for the development of circuit quantum electrodynamics~\cite{Burkard2020Mar}. At the heart of these devices lies the act of hybridizing the different super-semi constituents, each with their respective superconducting correlations and electrostatic charging effects leading to gate-tunable proximity effect~\cite{Chang2015Mar, Pendharkar2021Apr, Kanne2021Jul, Kousar2022Nov} and engineering of sub-gap states~\cite{Albrecht2016Mar, Kurtossy2021Oct, EstradaSaldana2022Apr, Dvir2023Feb}.



From the modelling perspective, one serious challenge in understanding such hybrids consists in combining
superconducting correlations with the potentially strong Coulomb interactions to be expected for confined geometries like semiconductor quantum dots and superconducting islands or grains. The sub-gap physics of quantum dots (QD) proximitized by grounded superconductors is well understood in terms of Yu-Shiba-Rusinov (YSR) states~\cite{Yu1965, Shiba1968, Rusinov1969, Baran2023Dec}. However, when also the superconductor has an electrostatic charging energy and constitutes a floating superconducting island (SI), the sub-gap physics may change considerably with states displaying at best a YSR-like character~\cite{EstradaSaldana2022Apr,Pavesic2021Dec}. 

The state-of-the-art approach for tackling SI-QD systems is based on the charge-conserving Richardson model~\cite{Pavesic2021Dec,Pavesic2022Feb,Saldana2022Feb,EstradaSaldana2022Apr,Zitko2022Jul, 
Bacsi2023Sep, Saldana2023Dec,Pavesic2024Feb}, inspired from earlier treatments of superconductivity in ultrasmall  grains~\cite{Dukelsky2000May}. In this context one deals with a number-conserving pairing Hamiltonian for the SIs (see Eq.~\ref{ham_SI} below) without resorting to the BCS mean-field approximation for the superconducting condensate. By preserving the particle number conservation, one may incorporate the charge fluctuations and account for the effects of its Coulomb repulsion effects. In the absence of Coulomb repulsion, one could safely use a BCS mean field description for SIs in present-day hybrid devices, and in fact a large $\mathcal{O}(100-1000)$ number of levels have been considered in the above works in the attempt to mimic the thermodynamic BCS limit. The high computational cost required to solve the full Richardson model even for simple devices may be traced back to the large amount of finite-size information that it carries. Naturally, this has prompted the search for cheaper alternatives, e.g. the flat-band approximation~\cite{Zitko2022Jul,Pavesic2024Feb} which may yield results in qualitative agreement with the full model in certain situations but is otherwise limited in applicability~\cite{Bacsi2023Sep}.

While the Richardson model is capable of addressing on the same footing all parameter regimes of simple QD-SI systems, other approaches are bound by severe compromises in this regard. For example, more complex quantum circuits with several superconducting leads have been recently investigated in Ref.~\onlinecite{Matute-Canadas2023Dec}, but only in the simplifying infinite-gap limit. Whereas the Cooper pair dynamics might be reasonably well accounted for, the ability to describe any YSR physics is completely lost as the normal regions are fully proximitized, and it is impossible to give any reliable predictions e.g. with respect to their response to quasiparticle poisoning. 
This infinite-gap modeling strategy is however computationally lightweight due to its complete disregard of any SI finite-size effects: a SI is taken into account only minimally through its proximity effect on a QD and through its $2e$ charge translations that serve to compensate the proximity-induced QD charge fluctuations, resulting in a simple number-conserving effective pairing Hamiltonian. While this way of restoring the particle-number conservation (by keeping track of the SI's charge translations) has been routinely used to account for the Coulomb repulsion in SIs that are free of finite-size effects~\cite{Fang2022Aug,Lapa2020Jun}, Ref.~\onlinecite{Pavesic2021Dec} argues that (a simple version of) this ``charge-counting trick~\cite{Lebanon2003Jul,Anders2004Nov} is not applicable to a gapped spectrum". 

In this work we construct an effective BCS mean-field description by stripping down the Richardson model of its finite size details while maintaining all $1e$ charging effects intact. This sets the stage for employing the few-level BCS surrogate model solver (SMS)~\cite{Baran2023Dec} to obtain a precise description of the QD-SI sub-gap physics. Finally, we restore the charge conservation by keeping track of the SI's charge translations in the presence of the surrogate model space (thus employing a more sensible version of the ``charge counting trick"). We thus end up with a general approach that is applicable in all parameter regimes while being free of finite-size effects and thus computationally highly efficient.

The remainder of this work is organized as follows: in Sec.~\ref{sec:method} we detail the construction of the generalized surrogate models which are then further developed and benchmarked against the Richardson results in Sec.~\ref{sec:benchmarks} and in the Appendices. In Sec.~\ref{sec:benchmarks} and ~\ref{sec:abs} we study several multi-QD systems using the surrogate models and comment on their physical interpretation that parallels the proximitized-nanowire setup. Finally, we draw conclusions in Sec.~\ref{sec:conclusions}.

\section{ Surrogate model methodology}
\label{sec:method}

\subsection{General modeling considerations}
\label{sec:IIA}

\begin{figure}[ht!]
\includegraphics[width=\columnwidth]{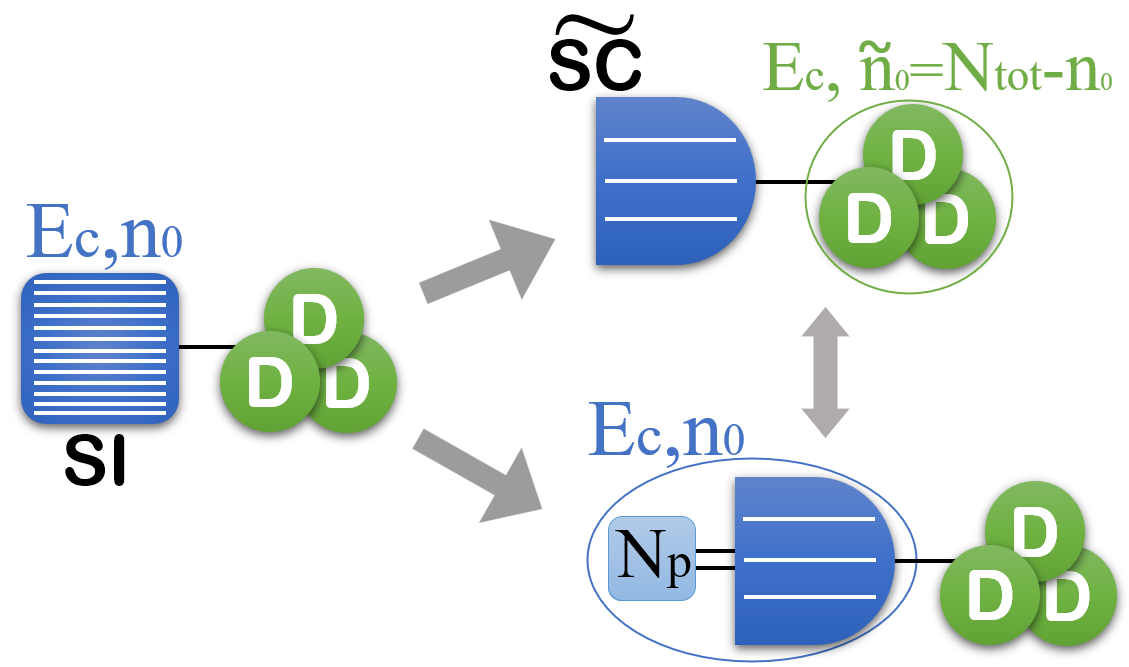}
\caption{Schematic of the generalized surrogate models. A superconducting island at strong pairing (blue, left) is coupled to multiple QDs (green dots), as detailed in Sec.~\ref{sec:IIA}. In the first approach (top right), the SI's charging term is transferred to the QDs via the particle number conservation, which then allows the SI to be approximated by a surrogate model, as in Sec.~\ref{sec:IIB}. In the second approach (bottom right), the SI is modelled as a surrogate model coupled to an auxiliary site counting the number of Cooper pairs $N_p$ in the condensate, as in Sec.~\ref{sec:IIC}.  \rev{The unitary equivalence between the two approaches is indicated by a double-headed arrow.} Single (double) lines indicate particle (pair) hopping.}
\label{fig_1}
\end{figure}

We first consider the simplest situation of one SI coupled to a relatively small system whose modelling can be performed efficiently without resorting to any further approximation, e.g. one or a few QDs as represented in Fig.~\ref{fig_1}. We consider the generic Hamiltonian
\begin{equation}\label{ham_general}
\begin{aligned}
\hat{H}=\hat{H}_{\text{SI},0}+E_c (\hat{N}_{\text{SI}}-n_0)^2+\hat{H}_{\text{QD}}+\hat{H}_{\text{tunn}}~,
    \end{aligned}
\end{equation}
where $E_c, N_{\text{SI}}$ and $n_0$ are the SI's charging energy, particle number and, respectively, the gate voltage
applied to the SI expressed in units of electron number. $\hat{H}_{\text{QD}}$ and $\hat{H}_{\text{tunn}}$ are the QD Hamiltonian (e.g., a collection of Anderson models) and, respectively, the SI-QD tunnel coupling.

The $\hat{H}_{\text{SI},0}$ term incorporates the superconducting correlations present in the SI. In Refs.~\cite{Pavesic2021Dec,Pavesic2022Feb,Saldana2022Feb,EstradaSaldana2022Apr,Zitko2022Jul, Malinowski2022Oct, Bacsi2023Sep, Saldana2023Dec}, in order to consistently account for the SI's considerable charging energy and the strong even-odd occupancy effects, it has been taken as the particle-number conserving Richardson model Hamiltonian, 
\begin{equation}\label{ham_SI}
H_{\text{SI},0}=\sum_{i=1}^L\sum_{\sigma=\uparrow \downarrow}\xi_i c^\dagger_{i\sigma} c_{i\sigma}-\lambda d \sum_{i,j=1}^L c^\dagger_{i\uparrow} c^\dagger_{i\downarrow} c_{j\downarrow} c_{j\uparrow}~,
\end{equation}
where $c^\dagger_{i\sigma}$ creates an electron with spin $\sigma$ and energy $\xi_{i}$ in the SI, $d=\xi_{i+1}-\xi_i$ is the level spacing (assumed constant) and $\lambda$ is the BCS coupling constant (e.g. for Al grains its value is $\lambda_{\text{Al}}=0.224$ \cite{Braun1998Nov}).

Depending on the ratio $d/\Delta=2\sinh (1/\lambda)/L$ (see Ref.~\onlinecite{Dukelsky2000May} and references therein),
between the level spacing $d$ and
the bulk superconducting gap $\Delta$, the Richardson Hamiltonian (\ref{ham_SI}) has two regimes.  On the one hand, the weak-coupling
 $d/\Delta\gg 1$ scenario, valid for small islands or small couplings $\lambda$, is characterized by strong pairing
fluctuations above the Fermi sea. On the other hand, the ground state physics in the strong-coupling regime $d/\Delta\ll 1$, valid for large islands (like the ones in contemporary
hybrid devices) or large couplings, is very well reproduced by the BCS wave function. In the latter regime one may formulate a surrogate model, provided the particle-number conservation is effectively taken into account as to correctly reproduce the charging effects.

\subsection{Surrogate models for a single SI coupled to QDs}
\label{sec:IIB}

For the purposes of computing the sub-gap spectrum of a sufficiently large SI with $d/\Delta\ll 1$, one should be able to formulate a model free of any finite size effects. Then, the low-energy physics would depend only on the relative difference between the total number of particles $N_{\text{tot}}$ and the SI's induced charge $n_0$: adding one Cooper pair ($N_{\text{tot}} \rightarrow N_{\text{tot}}+2$) while also adjusting $n_0$ accordingly ($n_0 \rightarrow n_0+2$) would have no observable effect.

To make this symmetry manifest, we use the conservation of the total number of particles for the SI - QD system,
\begin{equation}\label{number_conservation}
\begin{aligned}
\hat{N}_{\text{tot}}=\hat{N}_{\text{SI}}+\hat{N}_{\text{QD}}~,
    \end{aligned}
\end{equation}
to transfer the charging term in Eq.~(\ref{ham_general}) off the SI and onto the QDs,
\begin{equation}\label{ham_transfer}
\begin{aligned}
\hat{H}=\hat{H}_{\text{SI},0}+E_c [\hat{N}_{\text{QD}}-(\hat{N}_{\text{tot}}-n_0)]^2+\hat{H}_{\text{QD}}+\hat{H}_{\text{tunn}}~.
    \end{aligned}
\end{equation}
The Hamiltonian displays explicitly the physically relevant difference 
\begin{equation}\label{n0}
\tilde{n}_0\equiv N_{\text{tot}}-n_0~,
\end{equation}
which can now be thought of as an independent quantity. \rev{With $\tilde{n}_0$ held fixed, one is then free to consider a coherent superposition of different numbers of Cooper pairs in the SI (given that their actual number is irrelevant in the strong coupling regime assumed for the SI), which may be schematically written in a basis of unnormalized particle-number states as
\begin{equation}
    |\text{BCS}\rangle=\sum_{k\in \mathbb{N}} |N_\text{SI}+2k; n_0+2k\rangle~.
\end{equation}
Importantly, the components of the BCS state are wavefunctions shifted in both the SI's particle number  $N_\text{SI}$ and in the SI induced charge $n_0$ by the same number of pairs $k$. Thus, we effectively adopt a standard BCS mean-field description. }
Despite the apparent breaking of the particle number conservation, the substitution $H_{\text{SI},0}\rightarrow H_{\text{BCS}}$ in Eq.~(\ref{ham_transfer}), with 
\begin{equation}\label{ham_BCS}
\hat{H}_{\text{BCS}}=\sum_{j}\sum_{\sigma=\uparrow \downarrow}\xi_j c^\dagger_{j\sigma} c_{j\sigma}- \sum_{j} (\Delta c^\dagger_{j\uparrow} c^\dagger_{j\downarrow}+\Delta c_{j\downarrow} c_{j\uparrow})~,
\end{equation}
and $\tilde{n}_0$ \rev{of Eq.}~(\ref{n0}) held fixed, leads to results which are in excellent agreement  with those of the full Richardson model regarding the sub-gap spectrum and QD observables of an SI-QD system in the strong coupling regime (up to finite size effects), as also shown numerically in the next Section.

Within the BCS approach, the results mentioned above may be efficiently obtained by the standard surrogate model solver methodology of Ref.~\onlinecite{Baran2023Dec}, briefly summarized below. In this context, the full BCS model is replaced by a few-level surrogate whose parameters are optimized as to best reproduce the full hybridization function with the QDs.

For the rest of this work, we assume a constant density of states, $\nu_F=1/(2D)$, in a band of half-width $D$ around the superconductor's Fermi surface, as well as energy-independent tunneling amplitudes $t$ to each QD. The lead degrees of freedom are readily integrated out to give rise to the following Nambu tunneling self-energy (hybridization function)~\cite{Bauer2007Nov, Meng2009Jun}: 
\begin{align}\label{Sigma}
    \Sigma_d^{\text{T}}(\omega_n)=-\Gamma
    \begin{pmatrix}
            i\omega_n & \Delta\\
            \Delta & i\omega_n 
     \end{pmatrix} g(\omega_n),
\end{align}
with Matsubara frequencies $\omega_n=(2n+1)\pi k_{\text{B}}T$ at temperature $T$, tunneling rate $\Gamma=\pi\nu_{F}|t|^{2}$, and the $g$ function defined as
\begin{align}\label{gcont}
g(\omega)&\equiv \frac{1}{\pi}\int_{-D}^D\text{d}\xi\,\frac{1}{\xi^2+\Delta^2+\omega^2}=\frac{2}{\pi}\frac{ \arctan\left(\frac{D}{\sqrt{\Delta^2+\omega^2}}\right)}{\sqrt{\Delta^2+\omega^2}}~.
\end{align}

In constructing the  simplest discrete effective bath that best reproduces the sub-gap states of the full model, we note that each level with energy $\xi$ contributes a factor of $(\xi^2+\Delta^2+\omega^2)^{-1}$ to the $g$ function (\ref{gcont}). We approximate the latter by combining only a small number of such factors,
\begin{equation}
\label{tildeg}
\begin{aligned}
    &\tilde{g}_{\text{even}}(\omega)\equiv2\sum_{\ell=1}^K  \frac{\gamma_\ell}{\tilde\xi_\ell^2+\Delta^2+\omega^2}~, ~\tilde L =2K~,\\
    &\tilde{g}_{\text{odd}}(\omega)\equiv\frac{\gamma_0}{\Delta^2+\omega^2}+\tilde{g}_{\text{even}}(\omega)~,~\tilde L =2K+1~ ,\\
    \end{aligned}
\end{equation}
where $K$ denotes the number of pairs of effective levels. Such a $\tilde g$ function may be obtained by integrating out an effective superconducting bath with the same gap, $\Delta$, as the original one and whose $\tilde L$ discrete levels with energies $\pm|\tilde\xi_\ell|$ are coupled to the QD via a tunneling matrix elements $\tilde t_\ell=\sqrt{\gamma_\ell \Gamma}$. Each odd-$\tilde L$ model involves one extra level at zero energy, $\tilde \xi_0=0$. The effective bath is thus defined by parameters $\{\gamma_\ell, \tilde \xi_\ell\}$. We refer to $\tilde{L}$ as the surrogate bandwidth, and note that the case of $\tilde{L}=1$ corresponds to the so-called zero-bandwidth (ZBW) model. 

\rev{The SMS approach detailed in this section is represented schematically in the top right of Fig.~\ref{fig_1}}, with the explicit SMS Hamiltonian for one SI coupled to $N$ quantum dots being given by 
\begin{align}\label{hamtilde}
\widetilde{H}&=\hat{H}_{\widetilde{\text{SC}}}+\sum_{\alpha=1}^N (\hat{H}_{\text{QD},\alpha}+\hat{H}_{\widetilde{\text{T}},\alpha})\\
&+E_c\left(\sum_{\alpha=1}^N \hat{N}_{\text{QD},\alpha}-\tilde{n}_0\right)^2,\nonumber\\
     \hat{H}_{\widetilde{\text{SC}}}&=\sum_{\ell=1}^{\tilde{L}} \sum_{\sigma=\uparrow \downarrow} \tilde\xi_{ \ell} c_{\ell \sigma}^{\dagger} c_{\ell \sigma}- \sum_{\ell=1}^{\tilde L}(\Delta c_{\ell\uparrow}^{\dagger} c_{\ell \downarrow}^{\dagger}+\text{h.c.}),\nonumber\\
     \hat{H}_{\text{QD},\alpha}&=U_\alpha(\hat{N}_{\text{QD},\alpha}-\nu_\alpha)^2~,\nonumber\\
         \hat{H}_{\widetilde{\text{T}},\alpha}&=\sum_{\ell=1}^{\tilde L}\sum_{ \sigma=\uparrow \downarrow}\sqrt{\gamma_\ell \Gamma_\alpha}( c_{ \ell \sigma}^{\dagger} d_{\alpha\sigma}+\text {h.c.})~.\nonumber
\end{align}
Besides renormalizing the individual charging parameters of each QD, the $E_c$ term is seen to introduce capacitive couplings between the various QDs. As a technical note, one should remember to match the conserved even/odd total fermion parity of the BCS model with the even/odd total particle number $N_\text{tot}$ of the original model that enters $\widetilde{H}$ defined above through $\tilde{n}_0=N_\text{tot}-n_0$.

\subsection{Surrogate models for systems with multiple SIs}
\label{sec:IIC}

The methodology introduced in the previous subsection is limited to the efficient description of only one of the SIs in the system, as the transfer of the charging term cannot be done again after adopting the BCS picture. It would thus be desirable to revert back to a particle-number conserving model for each SI, while still retaining the computational benefits of the BCS-based surrogate models.

 To achieve this, we introduce an auxiliary degree of freedom comprised of the canonically conjugate number  and phase operators $\hat{N}_p$ and $\hat{\phi}$, $[\hat{N}_p,e^{i\hat{\phi}}]=e^{i\hat{\phi}}$. Physically, $\hat{N}_p$ counts the number of Cooper pairs in the superconducting condensate, while $e^{\pm i\hat{\phi}}$ adds/removes one pair from the condensate. The auxiliary Hilbert space is spanned by states $|p\rangle$ with an integer number of pairs $p\in \mathbb{Z}$, obeying $\hat{N}_p|p\rangle=p|p\rangle$ and $e^{\pm i\hat{\phi}}|p\rangle=|p\pm 1\rangle$.

We use this new degree of freedom in the surrogate Hamiltonian of Eq.~(\ref{hamtilde}) as a bookkeeping device for the number of Cooper pairs,
\begin{equation}\label{ham_SC_aux}
    \hat{H}^{\text{(aux)}}_{\widetilde{\text{SC}}}=\sum_{\ell=1}^{\tilde{L}} \sum_{\sigma=\uparrow \downarrow} \tilde\xi_{ \ell} c_{\ell \sigma}^{\dagger} c_{\ell \sigma}- \sum_{\ell=1}^{\tilde L}(\Delta c_{\ell\uparrow}^{\dagger} c_{\ell \downarrow}^{\dagger} e^{-i\hat{\phi}}+\text{h.c.})~,
\end{equation}
the resulting total Hamiltonian being  equivalent with the original one in Eq.~(\ref{hamtilde}) under the unitary transformation $c_{\ell\sigma}\rightarrow c_{\ell\sigma}\, e^{i\hat{\phi}/2}$.

The role of the superconducting Hamiltonian (\ref{ham_SC_aux}) is to conserve the particle number to which both the surrogate, and the auxiliary (Cooper pair condensate) Hilbert spaces contribute,
\begin{equation}
   \hat{N}^{\text{(aux)}}_{\widetilde{\text{SC}}}=\sum_{\ell=1}^{\tilde{L}} \sum_{\sigma=\uparrow \downarrow} c_{\ell \sigma}^{\dagger} c_{\ell \sigma}+2\hat{N}_p~.
\end{equation}

With the resurfaced $U(1)$ conserved charge at hand, we may now retrace our steps from the previous subsection, and arrive at
\begin{equation}
\label{htildetilde}
\widetilde{\widetilde{H}}= \hat{H}^{\text{(aux)}}_{\widetilde{\text{SC}}}+E_c[\hat{N}^{\text{(aux)}}_{\widetilde{\text{SC}}}-n_0]^2+\sum_{\alpha=1}^N (\hat{H}_{\text{QD},\alpha}+\hat{H}_{\widetilde{\text{T}},\alpha})~.
\end{equation}

The net result of this entire procedure is the effective replacement 
\begin{equation}\label{replacement}
   \hat{H}_{\text{SI},0}+E_c (\hat{N}_{\text{SI}}-n_0)^2 \rightarrow \hat{H}^{\text{(aux)}}_{\widetilde{\text{SC}}}+E_c[\hat{N}^{\text{(aux)}}_{\widetilde{\text{SC}}}-n_0]^2
\end{equation}
in the original Hamiltonian of Eq.~(\ref{ham_general}), where no assumptions about the rest of the system were made, other than the total particle number conservation. \rev{The SMS approach detailed in this section is represented schematically in the bottom right of Fig.~\ref{fig_1}}. It thus    becomes possible to build surrogate models for systems with an arbitrary number of superconducting islands by applying Eq.~(\ref{replacement}) to each one.

\section{Benchmarks and applications}
\label{sec:benchmarks}

\begin{figure}[ht!]
\includegraphics[width=\columnwidth]{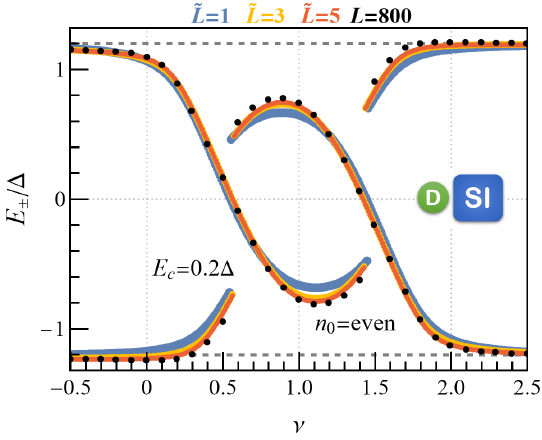}
\caption{QD-SI sub-gap excitation spectrum as a function of the gate voltage applied on the QD, as given by the $\tilde{L}=1,3,5$ surrogates ($\omega_c=10\Delta$). The positions $E_+ = E_1 - E_0$ and $E_- = E_0-E_{-1}$ are the excitation
energies of particle-like (+1) and hole-like (-1) states. The $L=800$ Richardson model results have been read graphically from Fig.~4 of Ref.~\onlinecite{Pavesic2021Dec} using WebPlotDigitizer \cite{Rohatgi2022}. The common parameters are $E_c=0.2\Delta$, $U=4\Delta, \Gamma=0.1U, \nu=1, D=40\Delta$ and $n_0$ an even integer.}
\label{fig_2}
\end{figure}

We solve the surrogate models introduced above by the density matrix renormalization group (DMRG) in the matrix-product-state formulation~\cite{White1992Nov, Schollwock2011Jan}, which is straightforward to implement with the ITensor library~\cite{itensor,itensor-r0.3}. Our numerical codes are available online~\cite{ github} and may be run on a standard laptop or desktop computer.

We show in Fig.~\ref{fig_2} an example of a sub-gap spectrum for the simplest QD-SI system. \rev{It was obtained by solving each of the $\tilde{L}=1,3,5$ surrogate models in various particle-number sectors, selecting the ground state energy energy $E_0$ (corresponding to, say, $N_0$ particles), and plotting the particle-like and hole-like excitation energies $E_\pm=\pm(E_{\pm 1}-E_0)$ of the states with $N_0\pm 1$ particles.} In all cases we find an excellent agreement with the Richardson model data of Ref.~\onlinecite{Pavesic2021Dec} even for a modest $\tilde{L}=3$ surrogate. All surrogates account well for the asymmetry in the sub-gap peak positions, which is a direct consequence of the Coulomb repulsion (at $E_c=0$ the spectra are the familiar symmetric eye-shaped loops, see e.g. Fig.~\ref{fig_8}a).

\begin{figure}[ht!]
\includegraphics[width=\columnwidth]{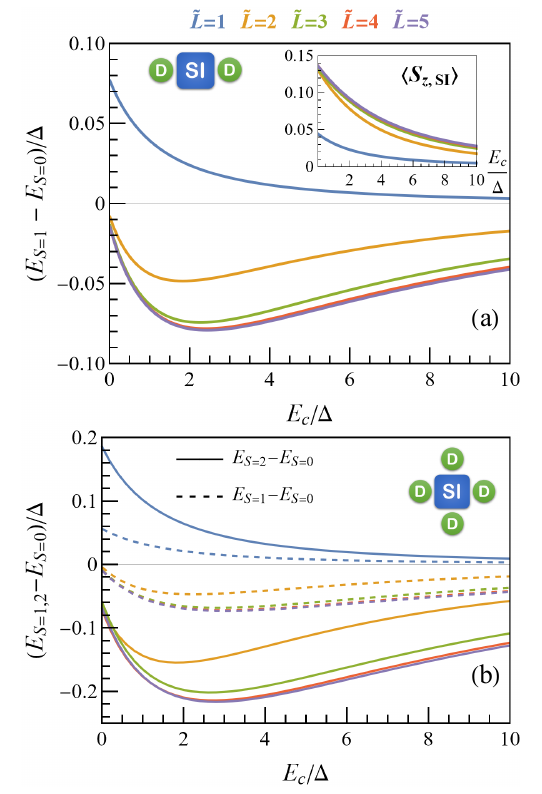}
\caption{(a) QD-SI-QD spin-singlet spin-triplet energy difference versus the SI's charging energy $E_c$, as obtained by the  $\tilde{L}=1,2,3,4,5$ surrogates at $U=6\Delta$, $\nu=1$, $\Gamma=\Delta$, with $n_0$ an even integer. Inset: average SI spin $\langle S_{z,\text{SI}}\rangle$ (in the $S_{\text{tot}}=S_{z,\text{tot}}=1$ state) versus $E_c$. (b) Same as in (a), but for the $S_\text{tot}=0,1,2$ states of a SI coupled to four identical QDs.}
\label{fig_3}
\end{figure}

\begin{figure}[ht!]
\includegraphics[width=\columnwidth]{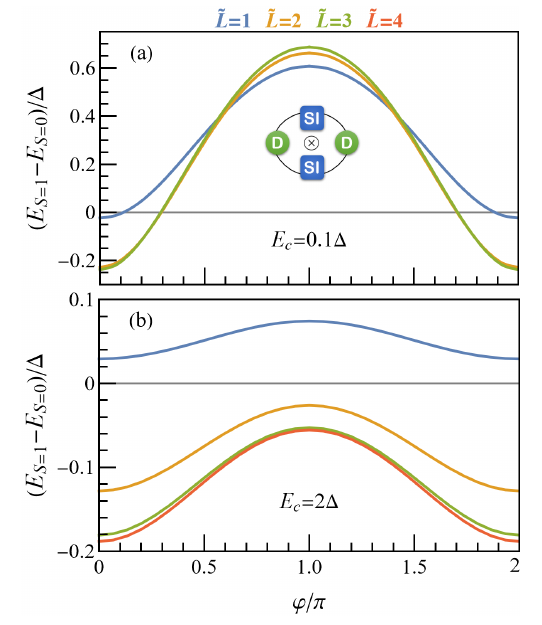}
\caption{ Spin-singlet spin-triplet energy gap for a QD-SI-QD-SI loop threaded by a flux $\varphi$, as obtained by the $\tilde{L}=1,2,3,4$ surrogates ($\omega_c=10\Delta$) for a small SI charging energy $E_c=0.1\Delta$ (a) and for a larger value $E_c=2\Delta$ (b).  The parameters are the same for all QDs and SIs, $ U=6\Delta,\nu=1, \Gamma=\Delta, D=10\Delta$, with $n_0$ chosen as an even integer.}
\label{fig_4}
\end{figure}

\rev{Additional benchmarks and further generalizations of the surrogate models are collected in the Appendices \ref{sec:app1},\ref{sec:app2} and \ref{sec:app3}. They include systems with multiple SIs, with combinations of SIs and grounded superconducting leads, and with direct coupling between SIs. Generally, we find a good qualitative agreement between all the surrogate and  Richardson model results.} Insofar as the considered system comprises a single QD (which may nevertheless be coupled to multiple SIs), all relevant physical aspects are qualitatively well accounted for even by the simplest $\tilde{L}=1$ (ZBW) surrogate. In this case, the finite bandwidth effects of the $\tilde{L}\geq 2$ surrogates only induce small to moderate quantitative differences in the sub-gap spectrum and related observables. The surrogate picture thus supports the adequacy of the flat-band approximation to the Richardson model \cite{Zitko2022Jul,Pavesic2024Feb} in such situations, but at the same time it offers a natural framework to safely model multiple-QD systems in which finite-bandwidth effects are known to be important~\cite{Bacsi2023Sep}.

For the rest of this Section, we focus on illustrative situations where non-trivial finite-bandwidth effects are at play, such for the QD-SI-QD setup and its generalizations. \rev{We address, from the SMS perspective, the subtle interplay between their spin-singlet and triplet ground states and explore how this is affected by the SI's charging effects}. The QD-SI-QD system has been recently studied in Ref.~\onlinecite{Bacsi2023Sep}, where it was found to undergo a spin-singlet to spin-triplet phase transition with increasing bandwidth. Let us first corroborate, within the SMS framework, the results of Ref.~\onlinecite{Bacsi2023Sep} regarding the basic properties of the QD-SI-QD system. Fig.~\ref{fig_2}a shows that the spin-triplet is indeed the ground state for all surrogate models except the $\tilde{L}=1$ (ZBW case), for an even SI occupancy $n_0$ and if $U$ is large enough such that each dot effectively hosts a single
electron ($\nu=1/2$).  The finite-bandwidth $\tilde{L}\geq 2$ surrogates are thus able to account for the basic mechanism that leads to the spin-triplet ground state, which involves breaking a Cooper pair and letting one of its members mediate the interdot ferromagnetic exchange \cite{Probst2016Oct, Bacsi2023Sep}.  While a large enough $E_c$  value eventually suppresses all charge fluctuations and with them the spin-singlet-triplet gap, at small $E_c$ the spin-singlet charge configuration is penalized by the SI's charging term in favor of the spin-triplet. For some intermediate $E_c$ value, the competition between these two effects leads to a robust maximum of the singlet-triplet gap, much larger than that corresponding to a grounded superconductor. Similar considerations apply to any number of dots, $N_\text{dots}$, individually and identically coupled to the same SI leading to an $S_{\text{tot}}=N_\text{dots}/2$ ferromagnetically aligned ground state, cf. Fig~\ref{fig_7}b.

The SMS approach allows us to easily and reliably assess the behaviour of even more complex tunable devices. We consider the QD-SI-QD-SI loop shown in Fig.~\ref{fig_4}a's inset which is threaded by a flux $\varphi$, treated as a control parameter. In the small $E_c$ limit, the general behavior may be understood in terms of QDs coupled to independent linear combinations of SC orbitals. At $\varphi=0$, both QDs effectively couple to the constructive superposition of the two SCs, while at $\varphi=\pi$ each QD forms a singlet with one of the even/odd combinations of SC orbitals (due to the relatively large value of the hybridization strength considered here). The formation of a spin-triplet state then requires breaking both singlet bonds, thus it is an excited state at $\varphi=\pi$ and small $E_c$, see Fig.~\ref{fig_4}a. The situation changes dramatically for a larger value of $E_c$, where the strong $E_c$-induced  coupling between the even and odd combinations of SC orbitals reinstates the spin-triplet as the global ground state of the $\tilde{L}\geq 2$ surrogates. As in the simpler QD-SI-QD setup, it is again crucial to take into account (at least in an minimal $\tilde{L}= 2$ way) the finite bandwidth effects for a sensible ground state description.

\section{Efficient SMS modeling of epitaxial super-semi nanowires}
\label{sec:abs}

\subsection{General considerations}

While for ultrasmall superconducting grains the finite-size effects play a significant role, the present-day hybrid superconducting islands (e.g. realized as epitaxial superconductor-semiconductor nanowires) show explicitly a dense continuum of states derived from the metallic superconducting shell above the bulk gap, and just a small number  $n^*$ of proximitized semiconductor sub-gap levels \cite{Saldana2023Dec}. Instead of using a brute-force discretization of the above-gap continuum as in the Richardson approach, we propose to model the proximitized nanowire in the more efficient SMS fashion.

The generalized SMS approach introduced in this work relies crucially on the existence of a dense continuum of superconducting levels. Given the absence of any finite-size details, we could still carefully break the $U(1)$ symmetry while keeping all $1e$ charging effects intact, which allowed us to build efficient BCS-based particle-number-conserving surrogate models (see Sec.~\ref{sec:method}). Our SMS approach contrasts with the Richardson methodology whose built-in finite-size effects can only be reduced by enlarging considerably the model space. The increase in computational complexity may be mitigated within the flat-band approximation, which is however not always applicable.

\begin{figure}[ht!]
\includegraphics[width=\columnwidth]{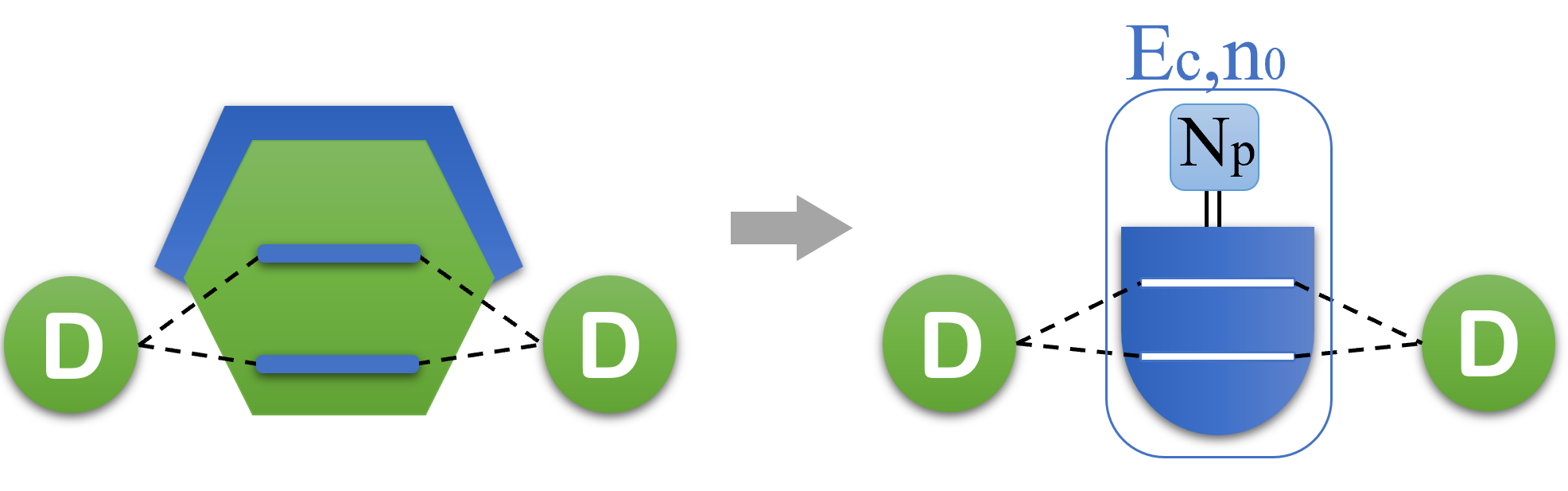}
\caption{A proximitized nanowire as a surrogate model: the metallic superconducting shell acts as a Cooper pair reservoir, while the proximitized semiconductor states are treated as effective surrogate  levels.}
\label{fig_5}
\end{figure}

In the absence of any charging effects, nanowire-base structures which also incorporate QDs may be modelled by tunnel-coupling the $n^*$ proximitized semiconductor levels (each with its induced BCS gap $\Delta^*_i$) to any relevant QD. Thinking of this as a surrogate model with $\tilde{L}=n^*$, we could invoke all the arguments detailed in Sec.~\ref{sec:method} to safely incorporate the charging effects, effectively ending up with a single extra auxiliary Cooper pair counter site in the model space, cf. Fig.~\ref{fig_5}. This  level structure is still much simpler than in the Richardson case, and it also lifts the need of a microscopic level-dependent effective pairing interaction, as used in Ref.~\onlinecite{Saldana2023Dec}. Instead, the only necessary model parameters are the values of the induced gap $\Delta^*_i$ for each level $i=1,...,n^*$. We find that the SMS-inspired approach is in excellent agreement with the results of Ref.~\onlinecite{Saldana2023Dec}, e.g. regarding the stepwise collapse of pairing correlations in an increasing external magnetic field. When going beyond this simplistic effective approach and towards more realistic scenarios, one could still employ a surrogate description of the metallic superconducting shell, e.g. if dealing with the details of the interface mediating the coupling of superconducting and semiconducting nanowire states.

\subsection{Applications}

\begin{figure}[ht!]
\includegraphics[width=\columnwidth]{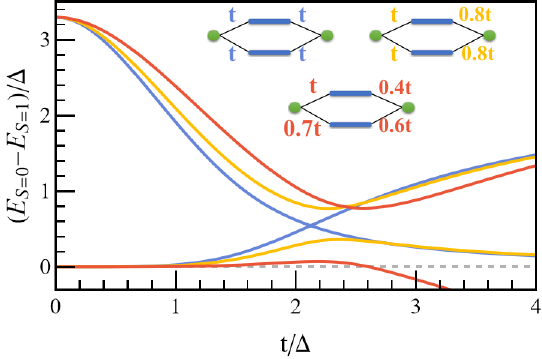}
\caption{QD-SI-QD spin-singlet-triplet energy gaps (for the two lowest-lying spin-singlets) versus the tunnel-coupling strengths for an $\tilde{L}=2$ surrogate at  $U=6\Delta$, $\nu=1$, $E_c=2\Delta$, with $n_0$ an even integer.}
\label{fig_6}
\end{figure}

\rev{Let us revisit in this context the QD-SI-QD setup. For simplicity, we assume for the SI a minimal model with $n^*=2$ sub-gap levels interpreted as proximitized semiconductor states, which are now allowed to couple differently to each QD. We also neglect any direct tunneling between the QDs, which would favor the spin-singlet ground state.}

The tunnel-coupling dependence of the spin-singlet-triplet gap shows in Fig.~\ref{fig_6} a competition between the two lowest-lying many-body spin-singlet states. In the vanishing limit of the QD-SI tunnel-couplings, the two QD spins may be trivially combined to $S_\text{tot} 	=0$  or $1$, the difference between the two lowest-lying spin-singlets (and spin-triplets) being the presence of a broken Cooper pair with an energy cost of $2E_\text{qp}$ (the members of the broken pair are free to form spin-singlets together and with the singly-occupied QDs). An increasing tunnel-coupling strength encourages the states with single QD occupation to hybridize with the SC quasiparticle excitations, while also allowing for empty/doubly-occupied dots. For a large enough tunneling strength, the lowest-lying spin-singlet inevitably acquires a strong broken-Cooper-pair character. This is manifest in Fig.~\ref{fig_6} where the two lowest-lying spin-singlet energies cross at some moderately large value of the coupling, in the left-right symmetric configuration. Breaking this mirror-symmetry leads to a spin-singlet avoided crossing and to a reduction of the maximum spin-singlet-triplet gap. In a highly asymmetric coupling situation, the broken-Cooper-pair-like singlet eventually overtakes the spin-triplet as the many-body ground state. The previous considerations apply in the convenient case of constructive interference across all paths in the system (when flipping the sign of one of the tunneling amplitudes, the parameter region with a triplet ground state is reduced).

\begin{figure}[ht!]
\includegraphics[width=\columnwidth]{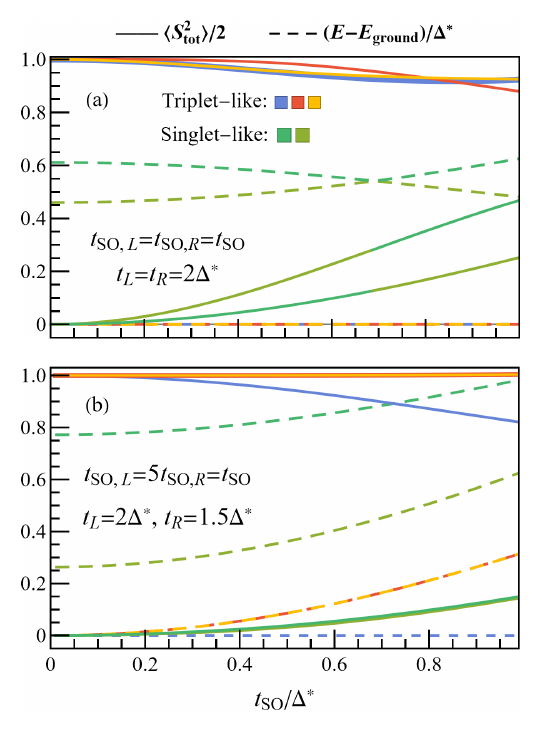}
\caption{ Total spin squared and excitation energy versus the spin-orbit tunneling strength $t_{SO}$ for the states of the system in (a).  Each QD$_{L,R}$ couples identically to both SC levels. The parameters are  $\Delta^*_{1,2}=\Delta^*$,  $\xi_{1,2}=\pm 1.3\Delta^*$, $E_c=2\Delta^*$, $t_\text{L}=t_\text{R}=2\Delta^*$, $t_{\text{SO,L}}=t_{\text{SO,R}}=t_{\text{SO}}$, $U=6\Delta^*,\nu=1$, with $n_0$ chosen as an even integer. (c) Same as in (b) but for an asymmetric setup $t_L=2\Delta^*$, $t_R=1.5\Delta^*$, $t_{\text{SO,L}}=5t_{\text{SO,R}}=t_{\text{SO}}$.}
\label{fig_7}
\end{figure}

We conclude this section by assessing the effects of spin-orbit interaction (SOI), which is strong in the InAs and InSb nanowires commonly employed in experiments \cite{Tosi2019Jan,Bommer2019May} and which we include here simply as a spin-flip tunneling term,
\begin{equation}
    \hat{H}_{\text{SO}}=\sum_{\alpha=1}^2\sum_{i=1}^{n^*} t_{\text{SO};\alpha, i} (-1)^\alpha  (d^\dagger_{\alpha\uparrow}c_{i\downarrow} -d^\dagger_{\alpha\downarrow}c_{i\uparrow} +\text{h.c.}) ~.
\end{equation}

Fig.~\ref{fig_7}a shows the evolution of the lowest-lying spin-triplet-like and spin-singlet-like states with increasing SOI strength $t_{\text{SO}}$ in the symmetric configuration $t_{\text{L}}=t_{\text{R}}=2\Delta^*$, $t_{\text{SO,L}}=t_{\text{SO,R}}=t_{\text{SO}}$. Here, the ground state retains a pronounced spin-triplet character up to relatively large values of $t_\text{SO}$, while the two  mid-gap excited singlet-like states are relatively stable in energy and experience a crossing around $t_\text{SO}/t \simeq 0.3$. Overall, the spin-triplet-like state experiences a much slower degradation of its total-spin character with increasing SOI strength than both spin-singlet-like excitations. In an asymmetrical coupling situation however, one of the spin-triplet components starts splitting in energy from the other two already around $t_\text{SO}/t \simeq 0.1$, while simultaneously and gradually losing its total-spin-1 character, cf. Fig.~\ref{fig_7}b. This suggests once again that a rather precise and controlled fabrication of the QD-SI-QD is necessary to achieve the strongly-coupled symmetric setup ideal for a robust spin-triplet unit.

\section{Conclusions}
\label{sec:conclusions}
In this work, we developed a class of charge-conserving few-level surrogate models for calculating efficiently the sub-gap spectrum of hybrid systems involving floating superconducting islands (SIs) and quantum dots (QDs).  We formulated the surrogate methodology around the basic assumption that the finite-size effects may be safely neglected, as is typically the case for present-day experimental implementation of SIs, e.g. in epitaxial semiconductor-superconductor nanowires. This enabled a BCS mean-field description of the charge-conserving hybrid system that a) perfectly accounts for all $1e$ charging effects and b) admits a highly efficient representation in terms of a very small number of surrogate effective levels \cite{Baran2023Dec}.

In all benchmarks our surrogates delivered essentially the same results (but free of finite-size effects) as the well-established Richardson model, while requiring only a minute fraction of its computational cost. Thus, we were led to argue that the surrogate approach is better suited for modelling the current generation of hybrid devices. Meanwhile, when looking from the other end of the modelling spectrum, our surrogate modelling strategy is free of the limitations that plague other computationally light-weight approaches such as the zero-bandwidth (flat-band) or infinite-gap approximations. We are confident that a surrogate-based study would provide valuable information on the physics of complex circuits based on multiterminal Josephson-Andreev junctions \cite{Matute-Canadas2023Dec}, e.g. regarding their robustness with
respect to quasiparticle poisoning.

Within the surrogate approach, we discussed the effects of the charging energy and spin-orbit interaction on the `exotic' triplet ground state recently predicted for QD-SI-QD based setups. Our main conclusion is that the spin-singlet-triplet gap is significantly enhanced by moderately large charging energy and hybridization strength, and is not affected much by the spin-orbit interaction in the symmetrically-coupled configuration. The triplet ground state is however found to be unstable with respect to disorder in the tunnel couplings. Given that extended QD-SI-QD chains can be fabricated with enough regularity, these findings raise the expectations regarding the physical implementation of a Heisenberg spin-1 chain in a super-semi hybrid platform. \rev{The theoretical investigation of long QD-SI-QD chains in the SMS framework is reported in Ref.~\cite{Baran2024Apr}.}

\begin{acknowledgments}

We thank R. \v{Z}itko and L. Pave\v{s}i\'{c} for valuable discussions. This work was supported by a grant of the Romanian Ministry of Education and Research, Project No. 760122/31.07.2023 within PNRR-III-C9-2022-I9.
\end{acknowledgments}

\appendix

\section{Additional SMS benchmarks}
\label{sec:app1}

We collect here additional benchmarks of the surrogate models against the Richardson solution of the QD-SI and SI-QD-SI systems. \rev{We limit ourselves here to discussing the way in which the SMS approach relates to the Richardson model, for in-depth physical discussions regarding the various systems see Refs.~\cite{Pavesic2021Dec,Pavesic2022Feb}.}

\begin{figure*}[ht!]
\includegraphics[width=\textwidth]{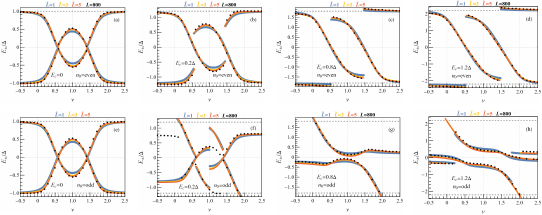}
\caption{Same as in Fig.~\ref{fig_2}, for even $n_0$ (a-d) and odd $n_0$ (e-h) and for various SI charging energies $E_c=0$ (a,e), $E_c=0.2\Delta$ (b,f), $E_c=0.8\Delta$ (c,g), $E_c=1.2\Delta$ (d,h).}
\label{fig_8}
\end{figure*}

\begin{figure*}[ht!]
\includegraphics[width=\textwidth]{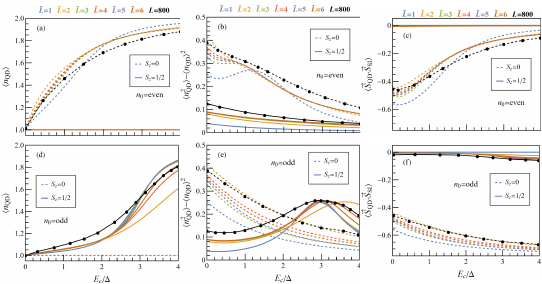}
\caption{QD-SI properties with increasing SI charging energy $E_c$: average QD occupancy (a,d), QD charge fluctuations (b,e) and QD-SI spin-spin correlation (c,f) in the lowest singlet (dashed curves) and doublet (continuous curves) at even $n_0$ (a,b,c) and odd $n_0$ (d,e,f), as given by the $\tilde{L}=1,...,6$ surrogates ($\omega_c=10\Delta$). The $L=800$ Richardson model results \rev{(black curves with dots)}  have been read graphically from Fig.~3 of Ref.~\onlinecite{Pavesic2021Dec} using WebPlotDigitizer \cite{Rohatgi2022}. The common parameters are $U=4\Delta, \Gamma=0.1U, \nu=1, D=40\Delta$. }
\label{fig_9}
\end{figure*}

\rev{Fig.~\ref{fig_8} shows the evolution of the QD-SI sub-gap spectrum with increasing SI charging energy $E_c$. From the familiar YSR eye-shaped loops at $E_c=0$, at large $E_c>\Delta$ the spectra tend towards straight lines typical of Coulomb blockaded systems}. The SMS spectra are by construction free of the finite-size effects inherent to the Richardson model (see e.g. Fig.~\ref{fig_8}f). Small quantitative differences between the surrogate and Richardson models are present also in the QD occupation, QD charge fluctuations and QD-SI spin-spin correlations, as shown in Fig.~\ref{fig_9}. In most cases such deviations persist all the way to $E_c=0$, which we interpret as being mainly related to the Richardson model's finite level spacing, $d=\Delta/10$ \cite{Pavesic2021Dec}. This is further supported by the good agreement (at $S_z=0$) between the QD charge fluctuations and the QD-SI spin-spin correlations for the Richardson model and the $\tilde L=2$ surrogate, featuring the largest level spacing amongst all surrogates. A detailed study of alternative SMS methodologies which better take into account the finite level spacing of small superconducting grains (e.g. by considering only even-$\tilde{L}$ cases with a minimum level spacing) is left for a future work.

\begin{figure}[ht!]
\includegraphics[width=\columnwidth]{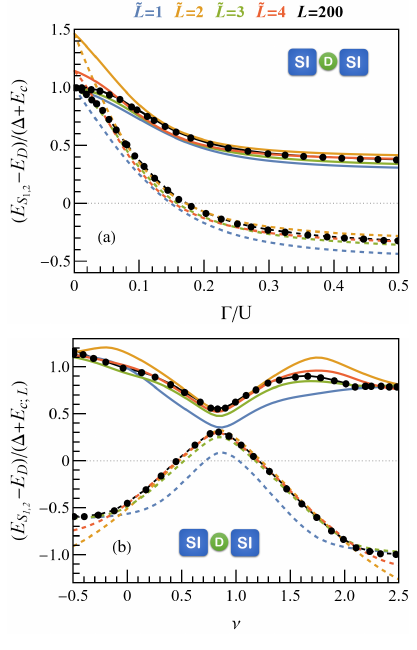}
\caption{SI-QD-SI sub-gap excitation spectrum (for particle addition) as a function of $\Gamma=\Gamma_L=\Gamma_R$ (a) and as a function of the gate voltage $\nu$ applied on the QD (b), as given by the $\tilde{L}=1,2,3,4$ surrogates ($\omega_c=10\Delta$). The energy of the lowest spin-doublet state \rev{$D$} is taken as the reference value, \rev{the curves indicating the lowest-lying spin singlets $S_{1,2}$}. The $L=200$ Richardson model results have been read graphically from Figs.~3 and 13b of Ref.~\onlinecite{Pavesic2022Feb} using WebPlotDigitizer \cite{Rohatgi2022}. The common parameters are $U=30\Delta, E_{c; \text{L}}=E_{c; \text{R}}=0.4\Delta, \nu=1, D=10\Delta$ for panel (a) and $U=4\Delta, E_{c; \text{L}}=0.1\Delta, E_{c; \text{R}}=1.5\Delta, \Gamma_L=0.1U, \Gamma_R=U, D=10\Delta$ for panel (b), with $n_0^{(\text{L})}=n_0^{(\text{R})}$ chosen as even integers. }
\label{fig_10}
\end{figure}

The SMS retains its fast convergence also for the more
complex SI-QD-SI system shown in Fig.~\ref{fig_10}. \rev{Here, we plot its subgap spectrum consisting of the lowest-lying spin-singlet states ($S_{1,2}$, with $n_0+2$ particles) and doublet state ($D$, with $n_0+1$ particles).}

\rev{The SI-QD-SI generalizes the SC-QD-SC setup discussed in detail in Ref.~\onlinecite{Baran2023Dec}), by including a non-vanishing charging energy for each superconducting island. For these systems, we generally find that odd-$\tilde{L}$ surrogates are better fitted for the weak tunnel coupling regime, whereas at strong coupling it is the even-$\tilde{L}$ models that show a faster convergence. At weak coupling, the even-$\tilde{L}$ surrogates are seen in Fig.~\ref{fig_10}a to slightly overestimate the excitation energy, as they lack the
screening quasiparticle with energy $\Delta$. At strong coupling, the odd-$\tilde{L}$ surrogates usually struggle to account for the interplay between the screening and localization effects
in the doublet state, see also the discussion around Fig. 3 in Ref.~\onlinecite{Baran2023Dec}. }

In Fig.~\ref{fig_10}b, all surrogates are shown to reproduce well the pronounced gate-dependence ($\nu$) of the sub-gap states, including the sweet-spot for qubit operation proposed in Ref.~\onlinecite{Pavesic2022Feb} for this SI$_\text{L}$-QD-SI$_\text{R}$ system. This two-level system is realized in the asymmetric configuration with SI$_\text{L}$ having a much larger charging energy and coupling strength to the QD than the SI$_\text{R}$ (which acts more as a particle reservoir, see Ref.~\onlinecite{Pavesic2022Feb} for more details).

\section{Surrogates models for systems with SIs and grounded superconducting terminals}

\label{sec:app2}

The number-conserving surrogates for floating SIs and number-violating surrogates for grounded BCS terminals may be seamlessly combined as building blocks of elaborate models for multi-terminal superconducting quantum circuits \cite{Matute-Canadas2023Dec}, capable of addressing sensitive issues such as their robustness with respect to quasiparticle poisoning.

\begin{figure}[ht!]
\includegraphics[width=\columnwidth]{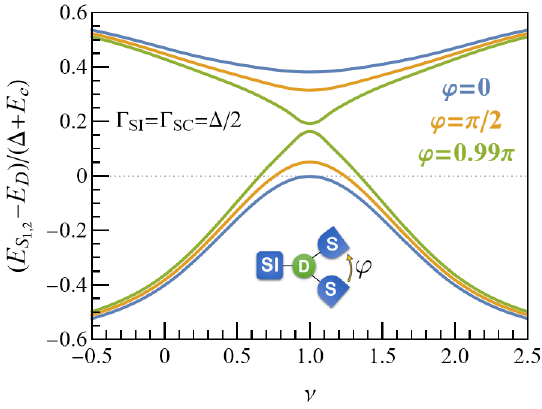}
\caption{SI-QD-SC$^2$ sub-gap excitation spectrum as a function of  the gate voltage $\nu$ applied on the QD for various superconducting phase differences $\varphi=0,\pi/2,\pi$, as given by the $\tilde{L}_{\text{SI}}=\tilde{L}_{\text{SC}^2}=2$ surrogates ($\omega_c=10\Delta$). The energy of the lowest spin-doublet state \rev{$D$} is taken as the reference value, \rev{the curves indicating the lowest-lying spin singlets $S_{1,2}$}.  The other parameters are $U=4\Delta, E_{c}=1.5\Delta, \Gamma_{\text{SI}}=\Gamma_{\text{SC},1}=\Gamma_{\text{SC},2}=0.5\Delta, D=10\Delta$, with $n_0$ chosen as an even integer. }
\label{fig_11}
\end{figure}

By combining the surrogate model of Eq.~(\ref{htildetilde}) for SI$_\text{L}$ with the original surrogate model of Ref.~\onlinecite{Baran2023Dec} for the right (grounded) SC terminal, we verified that the qubit sweet-spot of the SI-QD-SI setup survives in the strictly $E_{c;\text{R}}=0$ limit. Furthermore, we found the sweet-spot to be present also in a more complex three-terminal SI-QD-SC$^2$ setup, which generalizes the SI-QD-SC configuration (and effectively reduces to it for a vanishing phase difference $\varphi$ between the two SCs). Fig.~\ref{fig_11} shows that the sweet-spot's characteristics may be tuned by varying $\varphi$, and that a similar profile to the SI-QD-SI one of Fig.~\ref{fig_10}b may be recovered in the moderate coupling regime of the SI-QD-SC$^2$ setup around $\varphi=\pi$.

\section{SMS modeling of a Josephson junction}

\label{sec:app3}

\begin{figure}[ht!]
\includegraphics[width=\columnwidth]{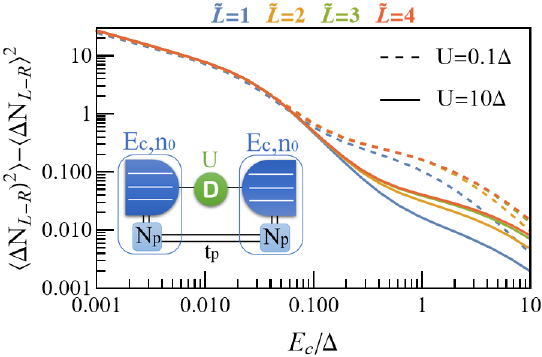}
\caption{Variance of the particle number difference $\Delta N_{\text{L-R}}=N_{\text{L}}-N_\text{R}$ versus the charging energy $E_c$, as given by the $\tilde{L}=1,2,3,4$ surrogates ($\omega_c=10\Delta$).  The other parameters are $\nu=0.9$, $ \Gamma_\text{L}=\Gamma_\text{R}=\Delta/4, t_p=0.1\Delta, \varphi_{\text{ext}}=\pi, D=10\Delta$, with $n_0$ chosen as an even integer. Compare with Fig.~6d of Ref.~\onlinecite{Pavesic2024Feb}.}
\label{fig_12}
\end{figure}

While the surrogate models considered above did account (by construction) for the pair-breaking tunneling processes induced by the presence of the  interacting QD, they may be easily generalized to describe also the dynamics of Cooper pairs, e.g. when modelling transmon qubits. 

Very recently, the full electronic problem for a combined Andreev spin qubit - transmon qubit system was solved in Ref.~\onlinecite{Pavesic2024Feb} within the flat-band approximation of the full SI$_\text{L}$-QD-SI$_\text{R}$ Richardson model. For the occasion, the latter was augmented by a pair-hopping term describing the Josephson junction with phase difference $\varphi_{\text{ext}}$, 
\begin{equation}
    H_{\text{JJ}}=t_p e^{i\varphi_{\text{ext}}}\frac{1}{L}\sum_{i=1}^L c^\dagger_{\text{L}i\uparrow} c^\dagger_{\text{L}i\downarrow}\sum_{i=1}^L c_{\text{R}j\downarrow}c_{\text{R}j\uparrow}+\text{h.c.}~,
\end{equation}
which may be effectively incorporated into the surrogate models as a coupling between the auxiliary Cooper pair condensate sites for each SI,
\begin{equation}
    H^{\text{(aux)}}_{\text{JJ}}=t_p\, e^{i \varphi_{\text{ext}}} \,e^{i\hat\phi_{\text{L}}}\,e^{-i\hat\phi_{\text{R}}}+\text{h.c.}~,
\end{equation}
see also the inset of Fig.~\ref{fig_12} for the graphical representation of the full system. \rev{Fig.~\ref{fig_12} shows the suppression of the charge fluctuations with increasing charging energy by plotting the variance of the particle number difference $\Delta N_{\text{L-R}}=N_{\text{L}}-N_\text{R}$ between the two SIs.} Our surrogate model results are shown there to be in excellent agreement with those of Ref.~\onlinecite{Pavesic2024Feb} (see Fig.~6b therein) regarding the two charge-fluctuations regimes observed when increasing $E_c$. The only quantitative differences between the various surrogates appear in the moderate to large $E_c$ regime where the $N_\text{L}=N_\text{R}$ state dominates and the electrostatic effects in the QD become relevant.

\bibliography{apssamp}

\end{document}